\documentclass[11pt,a4paper]{article}

\usepackage[utf8]{inputenc}
\usepackage[T1]{fontenc}

\usepackage{amsmath,amssymb}

\usepackage[hidelinks]{hyperref}
\usepackage{xurl}  

\usepackage[margin=2.5cm]{geometry}
\usepackage{setspace}
\usepackage{titling}
\usepackage{authblk}
\usepackage{graphicx}
\usepackage{booktabs}
\usepackage{array}
\usepackage{tabularx}
\usepackage{caption}
\usepackage[numbers,sort&compress]{natbib}

\usepackage{xcolor}
\usepackage{enumitem}
\setlist{itemsep=2pt, topsep=4pt}

\providecommand{\keywords}[1]{\vspace{0.5em}\noindent\textit{Keywords---} #1}

\newcommand{\prim}{\textbf{primary}}
\newcommand{\seco}{secondary}
\newcommand{\nap}{---}

\title{Bridging the Cybersecurity Gap Between Web2 and Web3\\
       \large An Incident-Based Analysis of Organizational and
              Application-Level Security Failures}

\author[]{Tarkan Yavas\thanks{M.Eng., CISM, ISO27001 Lead Auditor.
                              \texttt{tarkan.yavas@chain-horizon.com}}
\and Arslan Br\"omme\thanks{Dipl.-Inform., B.Sc., CISSP, CISA, CISM.
                            \texttt{arslanb@chain-horizon.com}}}
\affil[]{Chain Horizon GmbH\\
         Kolonnenstra\ss{}e 8, 10827 Berlin, Germany}

\date{Version 0.9.9.8}

\begin{document}

\nocite{ref01,ref02,ref03,ref04,ref05,ref06,ref07,ref08,ref09,ref10,
        ref11,ref12,ref13,ref14,ref15,ref16,ref17,ref18,ref19,ref20,ref21}

\maketitle

\begin{center}
\fbox{\parbox{0.92\linewidth}{\centering\small\itshape
Preprint / working paper. This version is a work in progress and may be
updated. Comments are welcome.}}
\end{center}
\vspace{0.5em}

\begin{abstract}
The rapid adoption of Web3 infrastructures has led to a growing number of
security incidents affecting cryptocurrency exchanges, custody services and
blockchain-based platforms. While existing research predominantly focuses on
vulnerabilities in smart contracts and blockchain protocols, a substantial
portion of real-world losses originates from off-chain systems,
organizational processes and human-centered operational workflows.

This paper presents a qualitative, incident-based analysis of publicly
documented, high-impact security breaches in the Web3 ecosystem, including
the Bybit exchange incident (2025), the Ronin Network bridge compromise
(2022), and the DMM Bitcoin exchange breach (2024). The selected cases are
systematically analysed and mapped to established Web2 security reference
frameworks, including OWASP-based vulnerability categories and organizational
security control domains.

The results indicate that dominant failure patterns in Web3 environments are
insufficiently addressed by generic security control catalogues, particularly
with respect to cryptographic key management, transaction approval
governance, signer and validator infrastructure, third-party tooling
dependencies, and human-in-the-loop processes.

Based on these findings, this paper argues for the adoption of established
information security management systems (ISMS) in Web3 organizations and
derives a structured set of blockchain-specific cybersecurity control
categories to operationalize existing ISMS frameworks for blockchain-based
systems. The proposed categories aim to bridge the gap between generic
security governance frameworks and domain-specific risks inherent to Web3
infrastructures.
\end{abstract}

\keywords{Web3 security, ISMS, OWASP Top 10, blockchain governance,
cryptocurrency exchanges, incident analysis, ISO/IEC 27001}

\section{Introduction}
\label{sec:intro}

The rapid growth of Web3 ecosystems has led to an increasing number of
organizations deploying blockchain-based infrastructures and decentralized
applications to manage digital assets, identity, and financial transactions.
While blockchain protocols such as Bitcoin and Ethereum provide strong
cryptographic guarantees at the protocol layer, a substantial share of
operational and security risk in Web3 environments arises outside the
blockchain itself. In particular, web-based applications, off-chain services,
organizational processes, and human--technology interactions constitute
critical parts of the overall attack surface.

In traditional Web2 environments, cybersecurity practices are commonly
guided by established international standards and regulatory frameworks,
such as ISO/IEC 27001, sector-specific compliance regimes, and structured
risk management processes. In many jurisdictions, organizations operating
critical digital infrastructures are required to undergo regular security
audits and to implement standardized control frameworks. By contrast,
Web3-oriented organizations frequently operate in highly dynamic and
innovation-driven environments in which comparable organizational and
technical security governance structures are not yet systematically
established.

Existing research and industry reports on Web3 security predominantly focus
on vulnerabilities at the smart contract and blockchain protocol layers.
However, recent high-impact incidents demonstrate that a significant
proportion of successful attacks exploit weaknesses in surrounding systems,
such as web applications, access and identity management components,
transaction approval workflows, and social engineering channels. These
components resemble classical Web2 architectures and are therefore exposed
to well-known categories of web application and organizational security
risks.

In Web2 environments, widely adopted reference frameworks such as the OWASP
Top 10 provide a structured taxonomy of common web application
vulnerabilities and serve as a practical baseline for secure software
development and testing. Nevertheless, empirical observations from Web3
practice indicate that foundational knowledge of established web application
security risks and corresponding mitigation strategies is not consistently
embedded in Web3 development and operational processes.

Beyond purely technical vulnerabilities, human-centered attack vectors,
including phishing, interface manipulation, and targeted social engineering,
represent an increasingly relevant threat to Web3 ecosystems. Due to the
irreversible nature of blockchain transactions, successful attacks can
result in immediate and unrecoverable financial losses, thereby amplifying
the impact of organizational and procedural security weaknesses when
compared to many traditional financial systems.

Despite the growing economic relevance of Web3 infrastructures, there is
currently limited empirical evidence on how far Web3 organizations adopt
established Web2 cybersecurity practices and to what extent recognized
security frameworks are reflected in their operational and development
processes. In particular, a systematic comparison between organizational
and application-level security practices in Web3 environments and
established Web2 reference models remains largely unexplored.

This study aims to address this gap by systematically examining publicly
documented, high-impact security incidents affecting Web3-oriented
organizations and by identifying recurring security weaknesses when
compared to widely recognized Web2 security frameworks.

The study is guided by the following research questions:

\begin{itemize}
\item[\textbf{RQ1:}] To what extent do major Web3 security incidents reflect
deficiencies in established Web2 cybersecurity practices, particularly with
respect to web application security and organizational security controls?
\item[\textbf{RQ2:}] Which categories of security controls, as defined by
established reference frameworks, are systematically underrepresented in
current Web3 operational environments?
\end{itemize}

This paper makes three main contributions:

\begin{enumerate}
\item It empirically characterizes how major Web3 security incidents
reflect deficiencies in established Web2 cybersecurity practices at both
the application and organizational levels.
\item It systematically maps incident-derived technical and organizational
weaknesses to OWASP and ISMS-oriented control domains, thereby providing
a structured lens for analysing off-chain and governance-related failure
patterns.
\item It derives and provides an empirically motivated prioritization of
blockchain-specific cybersecurity control categories that operationalize
existing ISMS frameworks for Web3 environments, offering a reusable
control structure for organizations operating blockchain-based
infrastructures.
\end{enumerate}

\section{Related Work}
\label{sec:related}

\subsection{Web2 Security Baselines for Application and Organizational
            Controls}

Established Web2 cybersecurity practice is anchored in internationally
recognized standards and reference frameworks. The OWASP Top 10 consolidates
consensus-driven categories of critical web application security risks and
is widely used as a baseline for secure development and testing
practices~\cite{ref01}.

On the organizational level, ISO/IEC 27001 defines requirements for
establishing, implementing, maintaining, and continually improving an
information security management system (ISMS) and is commonly applied as
a baseline for information security governance in both public and private
organizations~\cite{ref02}.

Complementary to ISO-based approaches, the NIST Cybersecurity Framework
(CSF) 2.0 provides a structured model for cybersecurity risk management
and governance and explicitly incorporates organizational, operational,
and supply-chain risk considerations~\cite{ref03}.

These baselines are directly relevant to Web3 environments because many
high-impact incidents originate in surrounding systems---web applications,
identity and access management, operational governance processes, and
third-party dependencies---rather than in blockchain cryptography or
consensus mechanisms.

\subsection{Web3 Security Research: Smart Contracts, DeFi and Bridge
            Infrastructures}

Academic research on Web3 security has predominantly focused on protocol-
and smart-contract-level vulnerabilities, particularly in decentralized
finance ecosystems. A comprehensive systematization of knowledge on
decentralized finance security organizes common attack mechanisms and
structural weaknesses inherent to composable smart-contract
architectures~\cite{ref04}.

Cross-chain interoperability has emerged as a particularly vulnerable
infrastructure layer. Recent academic work examines security challenges
of cross-chain bridges and highlights recurring design-level
weaknesses~\cite{ref05}. Complementary systematization research
categorizes major cross-chain bridge hacks and discusses recurring
vulnerability patterns and mitigation strategies~\cite{ref06}.

This body of literature provides a detailed understanding of on-chain and
protocol-centric risks. However, it largely abstracts from organizational
and operational environments in which blockchain systems are deployed,
operated, and governed.

\subsection{Exchange-Centric Incidents and Incident Datasets}

Centralized and decentralized exchanges represent a dominant operational
interface between users and blockchain systems. Academic analyses of
exchange architectures highlight fundamental differences in governance
models, custody structures, and operational responsibilities between
centralized and decentralized platforms~\cite{ref07}.

From an empirical perspective, incident-based research is supported by
recently published empirical studies and curated datasets that aggregate
historical cryptocurrency exchange incidents and associated financial
losses~\cite{ref08}. Industry-driven longitudinal analyses
document trends in stolen funds, incident volumes and evolving compromise
techniques~\cite{ref12}.

\subsection{Human-Centered Threats and User-Facing Attack Vectors}

Several studies emphasize that attackers frequently exploit the human
interface to security-critical operations rather than technical
vulnerabilities in blockchain protocols. User-centered threat models
explicitly cover phishing, wallet compromise, and deceptive transaction
approval mechanisms~\cite{ref09}.

A recent systematic review of Ethereum phishing detection research further
highlights the prevalence of scam- and social-engineering-driven attacks
and summarizes detection and mitigation approaches developed for blockchain
ecosystems~\cite{ref10}.

\subsection{Law-Enforcement and Public Attribution Sources}

Public advisories issued by law-enforcement agencies can serve as
high-quality primary sources for attribution and high-level attack
characterization. Such advisories provide independently verified context
for incident-based analyses relying on publicly available
information~\cite{ref11}.

\section{Methodology}
\label{sec:method}

This study adopts a qualitative, incident-based analysis of publicly
documented cybersecurity incidents affecting Web3-oriented organizations.
The objective is to identify recurring technical and organizational
weaknesses and to systematically relate these weaknesses to established
Web2 security reference frameworks.

The methodological design follows an exploratory and descriptive research
approach and is intended to support structured comparison across
heterogeneous real-world incidents.

\subsection{Data Sources and Case Selection}

The dataset consists exclusively of publicly available incident
documentation, including technical post-mortems, official statements by
affected organizations, reports by cybersecurity companies, investigative
journalism, and law-enforcement advisories.

Incidents were selected according to the following criteria:

\begin{itemize}
\item the affected organization primarily operates within the Web3
ecosystem, including cryptocurrency exchanges, custodial service providers,
blockchain infrastructures, or decentralized application platforms;
\item the incident resulted in a substantial and publicly reported
financial loss or severe operational disruption;
\item sufficient technical and organizational context was publicly available
to allow reconstruction of the primary attack vector and relevant enabling
conditions.
\end{itemize}

To ensure consistency with the research objectives, the study deliberately
focused on incidents whose root causes extended beyond blockchain protocol
vulnerabilities or smart contract logic flaws. Only incidents involving
off-chain components, operational workflows, governance mechanisms, or
human-centered processes were included.

Based on these criteria, three high-impact incidents were selected as
representative cases for in-depth qualitative analysis: the Bybit exchange
incident (2025), the Ronin Network bridge compromise (2022), and the DMM
Bitcoin exchange breach (2024). The selected incidents represent different
classes of Web3 organizations, including a centralized exchange, a
cross-chain bridge infrastructure and a custodial service provider.

\subsection{Analytical Framework}

To enable a structured comparison with established Web2 cybersecurity
practices, two complementary analytical perspectives were applied.

First, technical weaknesses and exploitation mechanisms were interpreted
using categories derived from established web application and system
security taxonomies, in particular vulnerability classes commonly reflected
in the OWASP Top 10 framework. This perspective supports classification of
failures related to authentication and authorization mechanisms, input and
transaction handling, interface manipulation, and system misconfiguration.

Second, organizational and operational weaknesses were analysed using
high-level information security management and governance domains derived
from widely adopted ISMS frameworks, including access governance, key and
credential management, third-party risk management, secure development
processes, operational monitoring, and incident response capabilities.

\subsection{Coding and Classification Procedure}

For each selected incident, the following analytical steps were performed:

\begin{enumerate}
\item identification of the primary attack vector as described in public
post-incident documentation and independent technical analyses
\item extraction of secondary enabling factors, such as insufficient
operational controls, missing governance mechanisms, inadequate
segregation of duties, or limited monitoring capabilities
\item classification of technical weaknesses according to established
vulnerability categories where applicable
\item classification of organizational and procedural weaknesses according
to high-level ISMS and governance control domains
\end{enumerate}

All extracted factors were documented and iteratively refined to ensure
consistency across cases. Classification and refinement were conducted by
both authors independently and subsequently reconciled through structured
discussion, applying a four-eyes principle to minimize confirmatory bias.
Where public information was ambiguous or incomplete, conservative
interpretations were applied and speculative assumptions were avoided.
The resulting classifications were subsequently aggregated to identify
recurring vulnerability patterns and dominant organizational control
deficiencies across the analysed incidents.

\subsection{Scope and Limitations}

This study relies exclusively on publicly available information. As a
consequence, the depth and precision of the analysis are constrained by
the level of detail disclosed in public incident reports and investigative
publications. Technical root causes may be incompletely reported, and
organizational failures may be underrepresented due to legal, reputational,
or regulatory considerations.

Furthermore, the selected cases represent high-impact incidents and do not
constitute a statistically representative sample of the entire Web3
ecosystem. Cases were selected using purposive sampling based on financial
impact, choosing the three highest-damage incidents in the Web3 space at
the time of writing. The convergence of all three cases on a single threat
actor profile (DPRK-attributed, state-sponsored) was not a selection
criterion but emerged as a finding, suggesting that the most severe Web3
security incidents to date share a common adversarial origin. This limits
generalizability to other threat actor profiles~-- such as opportunistic
attackers or malicious insiders~-- but reflects the current empirical reality
of high-impact Web3 attacks. The study therefore does not aim to quantify
the prevalence of specific vulnerabilities or control failures. Instead,
it focuses on identifying structural patterns and recurring weaknesses
observable in publicly documented real-world incidents.

Despite these limitations, incident-based qualitative analysis represents
an established approach in cybersecurity research and provides valuable
insights into real attack mechanisms and organizational failures,
particularly in rapidly evolving technological domains where large-scale
standardized datasets are not yet available.

Consequently, the study aims at analytic rather than statistical
generalization: the findings are intended to inform theory and control
design for Web3 security governance, not to estimate incident frequencies
or prevalence rates across the broader ecosystem.

\section{Results --- Selected High-Impact Incidents}
\label{sec:results}

This section presents the results of the incident-based analysis for the
three selected high-impact Web3 security incidents. Each case is described
with respect to the primary attack vector and the dominant organizational
and procedural enabling conditions, followed by an indicative mapping to
the applied reference frameworks.

\subsection{Bybit Exchange Incident (2025)}

\noindent\textbf{Date}\quad 21 February 2025\\
\textbf{Loss}\quad $\sim$\$1.5 billion USD\\
\textbf{Vector}\quad Interface manipulation and social engineering via
compromised third-party wallet frontend (Safe\{Wallet\})
\medskip

On 21 February 2025, the cryptocurrency exchange Bybit reported unauthorized
activity involving its Ethereum cold-wallet
infrastructure~\cite{ref15}. Law-enforcement authorities
reported that the incident resulted in the theft of approximately
\$1.5 billion USD in virtual assets and attributed the attack to
state-linked threat actors~\cite{ref11}.

The attack combined interface manipulation and social engineering techniques
that caused legitimate signers to approve malicious transactions. As a
result, valid transaction signatures were produced and subsequently executed
on-chain. The cryptographic integrity of the underlying blockchain network
remained intact throughout the incident.

From an organizational perspective, the incident revealed weaknesses in
governance and assurance processes for third-party wallet and signing
components, as well as insufficient independent verification mechanisms for
transaction context prior to approval. Furthermore, the absence of
real-time behavioural monitoring for signing workflows limited the
organization's ability to detect anomalous approval patterns in a timely
manner.

\subsubsection*{Indicative Mapping to Reference Frameworks}
The primary technical weakness in the Bybit incident corresponds to
A08~--~Software and Data Integrity Failures (OWASP Top 10:2021). The attack
exploited a compromised third-party signing frontend (Safe\{Wallet\}), which
delivered manipulated transaction data to the signing infrastructure without
being detected by integrity controls. This constitutes a supply-chain attack
on a critical software dependency. The reliance on an externally sourced
wallet frontend without adequate integrity assurance further maps to
A06~--~Vulnerable and Outdated Components.

A second primary weakness category is A04~--~Insecure Design: the
transaction approval workflow did not provide signers with an independent
means of verifying actual transaction content prior to signing. The absence
of real-time behavioural monitoring for signing workflows further
corresponds to A09~--~Security Logging and Monitoring Failures, which
precluded timely detection of anomalous approval patterns despite the
exceptional transaction volume involved.

At the organizational level, insufficient separation of duties within the
transaction approval process can be interpreted as reflecting
A01~--~Broken Access Control at the organizational workflow level. The
failure to authenticate the integrity of the presented transaction
information corresponds to A07~--~Identification and Authentication
Failures.

With respect to the proposed blockchain-specific control catalogue, the
incident most directly implicates BCMS-21 (Security of third-party service
integration), BCMS-07 (Security awareness and training), BCMS-05 (Audit
and monitoring systems for blockchain transactions), BCMS-17 and BCMS-20
(software update integrity and verification), BCMS-16 (Access controls for
blockchain administrators), and BCMS-19 (Secure design of blockchain
architectures). BCMS-01 and BCMS-22 apply as underlying governance control
gaps.

\subsection{Ronin Network Bridge Incident (2022)}

\noindent\textbf{Date}\quad March 2022\\
\textbf{Loss}\quad $\sim$\$615 million USD (173{,}600 ETH + 25.5M USDC)\\
\textbf{Vector}\quad Compromise of validator private key material and
operational control; insufficient segregation of validator infrastructure
(attributed to Democratic People's Republic of Korea (DPRK))
\medskip

In March 2022, the Ronin Network---an infrastructure supporting the Axie
Infinity ecosystem---suffered a major security breach in which approximately
173{,}600 ETH and 25.5 million USDC were drained from the Ronin bridge,
corresponding to an estimated loss of approximately 615 million
USD~\cite{ref13}.

In its official post-incident post-mortem, the operator reported that
attacker access to validator infrastructure enabled unauthorized
withdrawals. The incident was therefore not driven by a weakness in
blockchain cryptographic primitives, but by compromise of key material and
operational control over validator nodes and signing
processes~\cite{ref13}.

From an organizational and governance perspective, the incident illustrates
how concentration of operational authority and insufficient protection of
signing infrastructure can undermine the security assumptions of distributed
authorization systems. Public statements by the U.S. Department of the
Treasury linked laundering activity associated with the Ronin incident to
DPRK cyber threat actors~\cite{ref16}. Independent
blockchain forensics further corroborated the linkage to North Korean cyber
operations~\cite{ref17}.

\subsubsection*{Indicative Mapping to Reference Frameworks}
The Ronin Network incident is primarily characterized by weaknesses
corresponding to A01~--~Broken Access Control (OWASP Top 10:2021). The
compromise of validator credentials enabling the attacker to act as a
legitimate validator maps directly to A07~--~Identification and
Authentication Failures, while insufficient hardening of validator systems
and absence of adequate network segmentation constitute A05~--~Security
Misconfiguration.

The concentration of validator key material in a manner that permitted
compromise through a single attack path reflects A04~--~Insecure Design at
the governance architecture level. Although the cryptographic algorithms
were correctly implemented, the key management practices did not adequately
protect private key material from exfiltration, which additionally
corresponds to A02~--~Cryptographic Failures in an operational sense.
Limited monitoring of validator activity maps to A09~--~Security Logging
and Monitoring Failures.

With respect to the proposed blockchain-specific control catalogue, the
incident most directly implicates BCMS-03 (Protection of validators and
network nodes), BCMS-16 (Access controls for blockchain administrators),
BCMS-04 (Cryptographic key management), BCMS-19 (Secure design of
blockchain architectures), and BCMS-22 (Blockchain governance and
compliance).

\subsection{DMM Bitcoin Exchange Incident (2024)}

\noindent\textbf{Date}\quad May 2024\\
\textbf{Loss}\quad $>$\$300 million USD ($\sim$4{,}500 BTC)\\
\textbf{Vector}\quad Compromise of private keys for custodial wallets;
failures in key lifecycle management and privileged access control
\medskip

In May 2024, the Japanese cryptocurrency exchange DMM Bitcoin experienced
a large-scale security incident in which approximately 4{,}500 BTC were
transferred without authorization, corresponding to an estimated loss
exceeding 300 million USD~\cite{ref14}.

Public statements by law-enforcement authorities indicate that the incident
was enabled by the compromise of private keys associated with custodial
infrastructure rather than vulnerabilities in blockchain
protocols~\cite{ref14}. The incident reflects failures in the
protection, storage, and operational handling of cryptographic keys used
to control high-value custodial wallets. The resulting unauthorized
transactions were technically valid and irreversible once executed on-chain.

From an organizational perspective, the incident demonstrates persistent
weaknesses in custodial key lifecycle management, access control for
privileged systems, and operational segregation of duties in centralized
service environments.

\subsubsection*{Indicative Mapping to Reference Frameworks}
The DMM Bitcoin incident is most directly characterized by A02~--~Cryptographic
Failures (OWASP Top 10:2021): the root cause lies in the inadequate
protection, storage, and operational handling of cryptographic private keys.
The unauthorized access to cryptographic key material further maps to
A01~--~Broken Access Control, as privileged systems responsible for
custodial key operations lacked adequate access restrictions and
segregation of duties.

Weaknesses in the overall key lifecycle indicate the absence of
security-by-design requirements, corresponding to A04~--~Insecure Design.
Insufficient hardening of privileged systems reflects A05~--~Security
Misconfiguration. The absence of strong authentication controls maps to
A07~--~Identification and Authentication Failures, while limited monitoring
capabilities correspond to A09~--~Security Logging and Monitoring Failures.

With respect to the proposed blockchain-specific control catalogue, the
incident most directly implicates BCMS-04 (Cryptographic key management),
BCMS-09 (Protection of user wallets), BCMS-16 (Access controls for
blockchain administrators), and BCMS-22 (Blockchain governance and
compliance).

\subsection{Cross-Case Summary}
\label{sec:cross-case}

The following tables consolidate the framework mappings across the three
analysed incidents and provide the empirical basis for the
blockchain-specific control categories derived in
Section~\ref{sec:bcms-controls}.

Table~\ref{tab:owasp-mapping} maps the identified technical weaknesses to
OWASP Top 10:2021 vulnerability categories. \textit{Primary} relevance
indicates that the category directly characterizes the main attack vector
or a core enabling condition; \textit{secondary} relevance indicates a
contributing or amplifying factor.

\begin{table}[ht]
\centering
\small
\caption{Mapping of analysed incidents to OWASP Top 10:2021 categories.}
\label{tab:owasp-mapping}
\begin{tabular}{lccc}
\toprule
\textbf{OWASP Category} & \textbf{Bybit 2025} & \textbf{Ronin 2022} &
\textbf{DMM Bitcoin 2024} \\
\midrule
A01 --- Broken Access Control                  & \seco & \prim & \prim \\
A02 --- Cryptographic Failures                 & \nap  & \seco & \prim \\
A03 --- Injection                              & \nap  & \nap  & \nap  \\
A04 --- Insecure Design                        & \prim & \prim & \seco \\
A05 --- Security Misconfiguration              & \seco & \prim & \seco \\
A06 --- Vulnerable \& Outdated Components      & \prim & \nap  & \nap  \\
A07 --- Identification \& Authentication
        Failures                               & \seco & \prim & \seco \\
A08 --- Software \& Data Integrity Failures    & \prim & \nap  & \nap  \\
A09 --- Security Logging \& Monitoring
        Failures                               & \prim & \seco & \seco \\
A10 --- Server-Side Request Forgery            & \nap  & \nap  & \nap  \\
\bottomrule
\end{tabular}
\end{table}

The results indicate that no single OWASP category is dominant across all
three incidents. A04 (Insecure Design) shows primary relevance in two of
the three cases, reflecting that architectural and governance design
decisions---rather than implementation-level flaws---constitute a recurring
enabling condition in high-impact Web3 security failures. A03 (Injection)
and A10 (Server-Side Request Forgery) were not applicable to any of the
analysed cases.

Table~\ref{tab:bcms-mapping} maps the identified organizational and
technical weaknesses to the proposed blockchain-specific cybersecurity
control categories (BCMS). The same relevance classification applies.

\begin{table}[ht]
\centering
\footnotesize
\caption{Mapping of analysed incidents to proposed blockchain-specific
cybersecurity control categories (BCMS).}
\label{tab:bcms-mapping}
\begin{tabularx}{\linewidth}{l X c c c}
\toprule
\textbf{BCMS} & \textbf{Description} & \textbf{Bybit 2025} &
\textbf{Ronin 2022} & \textbf{DMM 2024} \\
\midrule
BCMS-01 & Blockchain-specific threat analysis and risk management
        & \prim & \prim & \prim \\
BCMS-02 & Security of consensus mechanisms
        & \nap  & \seco & \nap  \\
BCMS-03 & Protection of validators and network nodes
        & \nap  & \prim & \nap  \\
BCMS-04 & Cryptographic key management
        & \seco & \prim & \prim \\
BCMS-05 & Audit and monitoring systems for blockchain transactions
        & \prim & \seco & \seco \\
BCMS-06 & Incident recovery planning for blockchain environments
        & \seco & \seco & \seco \\
BCMS-07 & Security awareness and training for blockchain environments
        & \prim & \nap  & \nap  \\
BCMS-08 & Secure coding for smart contracts
        & \nap  & \nap  & \nap  \\
BCMS-09 & Protection of user wallets
        & \prim & \nap  & \prim \\
BCMS-10 & Security assessment of blockchain implementations
        & \seco & \seco & \seco \\
BCMS-11 & Encryption of blockchain-related data
        & \nap  & \seco & \seco \\
BCMS-12 & Monitoring of smart contracts
        & \seco & \nap  & \nap  \\
BCMS-13 & Security policies for off-chain data
        & \prim & \seco & \seco \\
BCMS-14 & Secure network protocols for blockchain communication
        & \nap  & \seco & \nap  \\
BCMS-15 & Incident response for blockchain-specific attacks
        & \seco & \seco & \seco \\
BCMS-16 & Access controls for blockchain administrators
        & \prim & \prim & \prim \\
BCMS-17 & Regular updates of blockchain software components
        & \prim & \nap  & \nap  \\
BCMS-18 & Use of zero-knowledge proofs for data protection
        & \nap  & \nap  & \nap  \\
BCMS-19 & Secure design of blockchain architectures
        & \prim & \prim & \seco \\
BCMS-20 & Validation and verification of blockchain updates
        & \prim & \nap  & \nap  \\
BCMS-21 & Security of third-party service integration
        & \prim & \nap  & \nap  \\
BCMS-22 & Blockchain governance and compliance
        & \prim & \prim & \prim \\
BCMS-23 & Protection against 51\% attacks
        & \nap  & \nap  & \nap  \\
BCMS-24 & Security assessments of smart contracts prior to deployment
        & \nap  & \nap  & \nap  \\
BCMS-25 & Secure storage of blockchain backups
        & \nap  & \seco & \seco \\
\bottomrule
\end{tabularx}
\end{table}

Three BCMS controls show primary relevance across all three incidents:
BCMS-01 (Blockchain-specific threat analysis and risk management), BCMS-16
(Access controls for blockchain administrators), and BCMS-22 (Blockchain
governance and compliance). This consistent pattern indicates that these
controls represent the most critical baseline requirements for Web3
organizations, irrespective of their specific operational model or primary
attack vector.

\section{Discussion}
\label{sec:discussion}

\subsection{Extension of the Threat Landscape: Physical Coercion Against
            Crypto Asset Holders}

While the incident analysis presented in this study focuses on technical
and organizational weaknesses in Web3 infrastructures, recent criminal
cases indicate the emergence of a complementary threat category that
cannot be addressed through conventional cybersecurity controls alone.

In particular, targeted physical coercion and extortion against individuals
controlling high-value cryptocurrency assets have been reported in multiple
jurisdictions. In 2025, the co-founder of a major hardware-wallet company
and his partner were abducted in France, physically assaulted and held for
ransom in cryptocurrency before being rescued by law-enforcement
authorities~\cite{ref18}. In a separate case in France,
law-enforcement authorities freed a victim following a kidnapping linked
to a wealthy cryptocurrency entrepreneur's family and ransom demands
involving digital assets~\cite{ref19}.

Similar incidents have also been reported outside Europe. In the United
States, a cryptocurrency investor was allegedly held captive and subjected
to physical violence in an attempt to coerce the disclosure of private
keys and access credentials to cryptocurrency wallets~\cite{ref20}.

In these scenarios, blockchain infrastructures and cryptographic mechanisms
remain uncompromised. Instead, the security failure occurs at the human
and physical security layer. This category of attacks differs fundamentally
from software- or infrastructure-based incidents such as those analysed in
this study. However, it directly exploits one of the defining
characteristics of decentralized systems: the absence of institutional
recovery mechanisms and the direct control of private keys by individuals.
Once coercion succeeds, the resulting transactions are technically valid
and irreversible.

From a risk management perspective, these incidents highlight the need to
extend security models for Web3 ecosystems beyond technical and
organizational controls towards integrated cyber-physical security
strategies. Such strategies include personal security awareness,
operational security practices for key holders, separation of key material
across individuals and locations, and organizational policies addressing
coercion and emergency response scenarios.

Consequently, physical coercion and extortion should be considered a
distinct and emerging threat category within the broader Web3 security
landscape, complementing the technical vulnerabilities and governance
failures identified in this study.

\subsection{Towards Blockchain-Specific Cybersecurity Controls for ISMS
            Frameworks}
\label{sec:bcms-controls}

While established information security management frameworks such as
ISO/IEC 27001 provide a robust and intentionally technology-agnostic
foundation for information security governance, the incident patterns
identified in this study indicate that several critical risk dimensions
of Web3 environments are insufficiently operationalized by generic
security control catalogues.

ISO/IEC 27001 focuses on management processes and high-level control
objectives and deliberately avoids domain-specific technical prescriptions.
However, the incidents analysed in this study demonstrate that Web3
environments introduce domain-specific assets, trust assumptions and
operational workflows---such as private key custody, transaction approval
processes, validator governance, wallet tooling and off-chain integration
layers---that require explicitly defined and consistently applied security
control categories.

Based on the recurring technical and organizational weaknesses observed in
the Bybit, Ronin Network and DMM Bitcoin incidents, this paper derives a
structured set of blockchain-specific cybersecurity control categories
that are intended to complement existing ISMS frameworks. The proposed
categories do not replace established ISMS controls, but operationalize
them for blockchain-based systems and custodial infrastructures.

The proposed control categories are based on the Blockchain Cybersecurity
Controls framework developed by the authors'
organization~\cite{ref21}. This framework was developed
independently of the present study as a generic control catalogue for
blockchain-based systems. The incident analysis conducted in this paper
does not generate the framework from scratch but uses the three
high-impact cases to empirically motivate, refine, and prioritize specific
control categories within it. The mapping logic applied here is not
limited to the authors' framework and can be transferred to alternative
blockchain security control catalogues.

\subsubsection*{Proposed Categories of Blockchain-Specific Cybersecurity
                Controls}

\paragraph{1.~Threat analysis and risk management}
\textbf{BCMS-01} --- Blockchain-specific threat analysis and risk
management
This category addresses the systematic identification, assessment and
treatment of blockchain-specific threat scenarios, including private key
compromise, transaction manipulation, signer infrastructure abuse,
validator collusion, compromised wallet tooling and cross-chain risk
propagation. The analysed incidents demonstrate that the absence of
structured, blockchain-specific risk modelling leads to an underestimation
of off-chain and governance-related risks.

\paragraph{2.~Security mechanisms and network protection}
\textbf{BCMS-02} --- Security of consensus mechanisms
\textbf{BCMS-03} --- Protection of validators and network nodes
\textbf{BCMS-14} --- Secure network protocols for blockchain communication
\textbf{BCMS-23} --- Protection against 51\% attacks
These controls address risks related to participation in consensus
mechanisms, protection of validator and node infrastructure, and the
integrity of peer-to-peer communication. The Ronin Network incident
illustrates how insufficient protection and governance of validator
infrastructure enables large-scale asset compromise without violating
cryptographic primitives.

\paragraph{3.~Key and data management}
\textbf{BCMS-04} --- Cryptographic key management
\textbf{BCMS-09} --- Protection of user wallets
\textbf{BCMS-11} --- Encryption of blockchain-related data
\textbf{BCMS-13} --- Security policies for off-chain data
This category operationalizes lifecycle management, storage, segregation
and controlled usage of private keys and sensitive off-chain data. Key
compromise and insufficient custody controls were the dominant enabling
factors in the DMM Bitcoin incident.

\paragraph{4.~Monitoring and auditability}
\textbf{BCMS-05} --- Audit and monitoring systems for blockchain
transactions
\textbf{BCMS-12} --- Monitoring of smart contracts
These controls focus on the detection of anomalous transaction patterns,
abnormal signing behaviour and deviations in smart contract execution.
The Bybit incident highlights the lack of operational monitoring for
transaction approval workflows as a critical detection gap.

\paragraph{5.~Emergency management and recovery}
\textbf{BCMS-06} --- Incident recovery planning for blockchain environments
\textbf{BCMS-25} --- Secure storage of blockchain backups
This category addresses preparedness for large-scale incidents, including
compromised keys, halted validator operations and partial loss of custodial
infrastructure. Due to the irreversibility of blockchain transactions,
recovery capabilities primarily focus on containment, service continuity,
controlled key rotation and infrastructure re-establishment.

\paragraph{6.~Training and awareness}
\textbf{BCMS-07} --- Security awareness and training for blockchain
environments
This category targets human-centred attack surfaces, including social
engineering, deceptive transaction approval and improper wallet handling.
The Bybit incident demonstrates how human-in-the-loop approval processes
become primary attack vectors in the absence of targeted training and
procedural safeguards.

\paragraph{7.~Secure development and deployment}
\textbf{BCMS-08} --- Secure coding for smart contracts
\textbf{BCMS-24} --- Security assessments of smart contracts prior to
deployment
Although the incidents analysed in this study were not primarily caused
by smart contract logic flaws, secure development and structured
pre-deployment assessments remain essential to reduce secondary
exploitation paths and systemic vulnerabilities.

\paragraph{8.~Governance and compliance}
\textbf{BCMS-21} --- Security of third-party service integration
\textbf{BCMS-22} --- Blockchain governance and compliance
This category addresses governance structures for validators, signers,
custody providers and external tooling vendors. The Bybit incident
illustrates how third-party wallet and signing components constitute
critical dependencies that require formal assurance, contractual controls
and continuous oversight.

\paragraph{9.~Access control and incident response}
\textbf{BCMS-15} --- Incident response for blockchain-specific attacks
\textbf{BCMS-16} --- Access controls for blockchain administrators
These controls focus on privileged access to signing infrastructure,
validator systems and custody environments. Across all analysed cases,
insufficient segregation of duties and limited oversight of privileged
operational access amplified the impact of technical compromise.

\paragraph{10.~Software and system updates}
\textbf{BCMS-17} --- Regular updates of blockchain software components
\textbf{BCMS-20} --- Validation and verification of blockchain updates
This category addresses update integrity and supply-chain risks for node
software, wallet infrastructure and signing services, thereby reducing
exposure to compromised dependencies and malicious updates.

\paragraph{11.~Innovative technologies and architectural safeguards}
\textbf{BCMS-18} --- Use of zero-knowledge proofs for data protection
\textbf{BCMS-19} --- Secure design of blockchain architectures
These controls support privacy-preserving and resilience-oriented
architectural patterns, including the reduction of sensitive data exposure
and improved separation of security-critical components.

\paragraph{12.~General blockchain security assurance}
\textbf{BCMS-10} --- Security assessment of blockchain implementations
This category provides a consolidated assurance layer for evaluating
blockchain platforms, custody infrastructures and operational architectures
against defined security baselines.

\medskip
Overall, the proposed blockchain-specific cybersecurity control categories
directly address the dominant failure patterns observed in the analysed
incidents, particularly in the areas of cryptographic key management,
transaction governance, signer and validator infrastructure, third-party
tooling and human-centred operational processes.

The mapping of the analysed incidents to the proposed control categories
further reveals a consistent cross-case pattern. Three controls---BCMS-01
(Blockchain-specific threat analysis and risk management), BCMS-16 (Access
controls for blockchain administrators), and BCMS-22 (Blockchain governance
and compliance)---show primary relevance across all three incidents,
irrespective of the specific operational model or attack vector involved.

Conversely, BCMS-08 (Secure coding for smart contracts), BCMS-18 (Use of
zero-knowledge proofs), and BCMS-23 (Protection against 51\% attacks) were
not directly implicated in any of the analysed cases, consistent with the
finding that dominant failure patterns originate outside the blockchain
protocol layer.

Integrating such domain-specific control catalogues into existing ISMS
frameworks can significantly improve the applicability of standardized
cybersecurity governance to Web3 environments and supports a transition
from isolated technical audits towards holistic organizational risk
management for blockchain-based systems.

\section{Conclusion}
\label{sec:conclusion}

This paper presented an incident-based analysis of three publicly
documented high-impact Web3 security incidents (Bybit 2025, Ronin 2022,
DMM Bitcoin 2024) and mapped recurring failure patterns to established
Web2 security reference frameworks and ISMS-oriented governance domains.

With respect to RQ1, the results indicate that major Web3 incidents
frequently reflect deficiencies in established Web2 cybersecurity
practices, particularly in areas such as secure authentication and
authorization workflows, privileged access governance, third-party
dependency assurance, and operational monitoring of high-risk processes.
The incidents analysed demonstrate that large-scale losses can occur
without compromising blockchain cryptographic primitives, as attackers
often succeed by targeting off-chain components, organizational
procedures, and human-in-the-loop approval mechanisms.

With respect to RQ2, the analysis highlights systematic underrepresentation
of domain-specific control requirements for Web3 environments in generic
security control catalogues. In particular, key lifecycle governance,
signing infrastructure protection, transaction approval governance,
validator and bridge operations, and wallet/tooling dependencies require
explicit and consistently applied control categories. To address this
gap, the paper proposed a structured set of blockchain-specific
cybersecurity control categories intended to complement existing ISMS
frameworks and support their operationalization in blockchain-based
systems.

Future work should evaluate the proposed control categories against
additional incident datasets, develop measurable control objectives and
maturity criteria, and validate their effectiveness through empirical
assessment in operational Web3 environments.

\section*{Availability of Control Catalogue Implementation}
\addcontentsline{toc}{section}{Availability of Control Catalogue Implementation}

An implementation of the proposed blockchain-specific cybersecurity control
categories is available as part of a governance, risk and compliance (GRC)
tool providing information security management system (ISMS) support and
governance capabilities, developed by the authors'
organization~\cite{ref21}.

The tool provides a structured representation of the control catalogue and
supports the mapping of organizational and technical security measures to
the proposed categories. The implementation is provided solely as a
practical reference and does not affect the methodological results or
conclusions of this study.

\section*{Use of AI Tools}
\addcontentsline{toc}{section}{Use of AI Tools}

The authors used the following AI-assisted tools during the preparation of
this manuscript: ChatGPT 5.2 (OpenAI), Claude Sonnet 4.5 (Anthropic,
accessed via Claude.ai), and Perplexity AI (Standard, Perplexity AI Inc.).
These tools were employed to support literature research, language editing,
structuring of the manuscript, and refinement of academic writing.

All intellectual contributions, including research design, analysis,
interpretation, and conclusions, remain solely those of the authors.
AI-generated outputs were reviewed, edited, and validated by the authors
prior to inclusion.

\section*{Competing Interests}
\addcontentsline{toc}{section}{Competing Interests}

Both authors are affiliated with an organization that develops a
governance, risk and compliance tool in which the proposed
blockchain-specific cybersecurity control categories are implemented.

The authors declare that this affiliation had no influence on the design,
analysis, or interpretation of the results presented in this study.

\bibliographystyle{unsrtnat}
\bibliography{references}

\end{document}